\documentclass[eqsecnum,preprint,prd,aps,nofootinbib]{revtex4}
\usepackage{epsfig}
\usepackage{amssymb}
\usepackage[dvips]{color}
\usepackage[latin1]{inputenc}
\usepackage{graphicx}

\newcommand{\nc}{\newcommand}
\nc{\be}{\begin{equation}} \nc{\ee}{\end{equation}}
\nc{\bea}{\begin{eqnarray}} \nc{\eea}{\end{eqnarray}}
\nc{\bt}{\begin{tabular}} \nc{\et}{\end{tabular}}
\nc{\ba}{\begin{array}} \nc{\ea}{\end{array}}
\nc{\dy}{\displaystyle} \nc{\pr}{{\rm I}} \nc{\se}{{\rm II}}
\nc{\w}{{\rm w}} \nc{\s}{{\rm s}} \nc{\um}{{1\over 2}}
\nc{\Hc}{{{\cal H}_{\rm c}}}

\def\s{\sigma}

\begin{document}
\begin{center}
\bibliographystyle{article}
{\Large \textsc{Non-commutative Kerr black hole}}
\end{center}

\date{\today}

\author{Elisabetta Di Grezia}
\email{digrezia@na.infn.it} 
\affiliation{Istituto Nazionale di Fisica Nucleare, Sezione di
Napoli, Complesso Universitario di Monte S. Angelo, Via Cintia
Edificio 6, 80126 Napoli, Italy}
\affiliation{Dipartimento di
Scienze Fisiche, Complesso Universitario di Monte S. Angelo, Via
Cintia Edificio 6, 80126 Napoli, Italy}
\author{Giampiero Esposito}
\email{giampiero.esposito@na.infn.it} 
\affiliation{Istituto Nazionale di
Fisica Nucleare, Sezione di Napoli, Complesso Universitario di
Monte S. Angelo, Via Cintia Edificio 6, 80126 Napoli, Italy}
\begin{abstract}
This paper applies the first-order Seiberg--Witten map to evaluate
the first-order non-commutative Kerr tetrad. The classical tetrad is
taken to follow the locally non-rotating frame prescription.  
We also evaluate the tiny effect of non-commutativity on the efficiency 
of the Penrose process of rotational energy extraction from a black hole.
\end{abstract}

\maketitle

\section{Introduction}

In general relativity the problems concerning the singularities in
gravitational collapse have led to important developments over decades
and are still at the core of much insight into Einstein's theory
\cite{chri09}.

In 1795 Laplace, relying upon Newtonian gravity, found that a very
dense and massive object would appear black, because light would
not be able to escape from it. Much later, in 1915, Einstein 
developed the theory of general relativity which predicts the
possible existence of such dark objects, black holes, caused by
singularities, which are objects with infinite curvature resulting
from an infinite density, so that everything nearby is drawn into the
black hole.

Soon after Einstein's seminal paper,
Schwarzschild obtained in 1916 the first solution of the Einstein
equation of general relativity, which described the space-time
around a static spherically symmetric massive object, that does
not have an angular momentum or charge. This was called, since
then, the Schwarzschild metric.

Later, in 1963, Kerr discovered another solution
\cite{kerr63}, the Kerr metric,
which describes the space-time outside a massive axi-symmetric
rotating object. A rotating black hole has rotation in addition to
the static black hole.

These two solutions describe the static and rotating black holes,
respectively. Thus, black hole solutions of the Einstein equations
are characterized by three parameters, i.e., mass $M$, angular
momentum $J$, and charge $Q$ by the no-hair, or uniqueness,
theorem \cite{wald84}.

The description of a rotating black hole uses two of the three
parameters: mass and angular momentum. The black hole mass, angular 
momentum and charge are conserved during the collapse of the
star, because of global conservation laws. All other properties of
the star collapsed to form the black hole are lost during the collapse
(no-hair theorem). Then there are four laws, derived from standard
laws of physics, which describe the thermodynamics of a black hole
\cite{hawk73}. In particular the temperature $T=\kappa /2\pi$ is the
Hawking temperature of the black hole, and is defined by
\cite{hawk73,nico05,nico06,anso07,spal09,nico09}:
\begin{equation}
T\equiv -\left(\frac{1}{4\pi\sqrt{-G_{00}G_{rr}}}
\frac{dG_{00}}{dr}\right)_ {r=r_H}.
\end{equation}
The Kerr metric describes the space-time
around a rotating or spinning singularity without charge
and time-independent, i.e. the axi-symmetric gravitational field of
a collapsed object that has retained its angular momentum. Stellar
black holes, on the other hand, 
are caused by the collapse of stars. A star is a very
massive, rotating but charge-less object. Because charges of
opposite sign cancel each other, stars are neutral. Hence, the
space-time around a stellar black hole is described by the Kerr
metric. Different coordinate systems are 
indeed used for the Kerr metric, and a coordinate
system commonly used is the Boyer--Lindquist 
\cite{boye67} which exploits
spherical coordinates. These are easy to work with and some
features are easily noticed, but it has a coordinate singularity
at the event horizon. Thus, by
using different coordinate systems, different properties or
features of a rotating black hole can be described.

In the second half of the twentieth century, the black hole is no
longer a theoretical construction, but it could be a real physical
object because astronomers observed very small objects that
emitted jets of particles with very high energy. They proposed
black holes as the sources of these jets 
\cite{dhav00,toms03}. In fact
massive stars undergoing a gravitational collapse in the final
state, are expected to become a black hole. Moreover, a rotating
star will collapse in a spinning or rotating black hole. Thus, the
Kerr metric is a good candidate to describe the space-time around
the final state of a very massive star.

Rotating black holes, discovered by Kerr as exact solutions of
general relativity equations, are of great astrophysical interest
because their emissions provide a method for identifying and
studying black holes.

There are several motivations for studying black holes and the
metrics that describe the space-time around them; one is that they
form objects which probably have to be understood in terms of
quantum gravity: large mass at small size. Second, because
stellar mass black holes give information about the {\em last
stage in star evolution}; third, these black holes are important
in cosmology since they could be {\em seeds of galaxy formation}.
However, it is unclear whether they are properly described by the
Kerr metric. Super-massive black holes can give information about
the {\em very early universe era} .

Hence there are three general types of black holes: stellar black
holes, primordial black holes and super-massive black holes,
distinguished by their mass and size. 

There are several ways in which black holes could be observed in
an indirect way: X-rays, spectral shift, gravitational lensing. The
only direct way to observe black holes is via gravitational waves.
So far, no black hole has been directly observed.

In astrophysical contexts $Q$ is negligible, because the electric
charge is shortened out by the surrounding plasma \cite{blan77}. Thus,
the variation in the observational properties of black holes is
due to external parameters, such as the angle between the black
hole spin vector and the line of sight, the gas accretion flow
geometry and black hole spin $j\equiv J/M^2=a/M$.

On choosing $G=c=1$ units, the Kerr metric in Boyer-Lindquist 
coordinates \cite{boye67} reads as (we actually follow
the sign conventions in Ref. \cite{bard72})
\begin{eqnarray}
ds^{2}&=& -\left({1-\displaystyle{2 M r \over \Sigma}}\right) 
dt^2 +{\Sigma \over
\Delta} dr^2 +\Sigma d\theta^2+\left\{(r^2+a^2)\sin^2\theta+{2M\,
a^{2}r\sin^4\theta \over \Sigma} \right \}d\varphi^{2} \nonumber \\
&-&{4M\,ar
\sin^2\theta \over \Sigma} dt d\varphi, 
\label{(1.2)}
\end{eqnarray} 
where $M$ is the mass of the black hole, $a$ is its angular momentum
per unit mass ($0 \leq a \leq M$),  
and the functions $\Delta$, $\Sigma$ are given by
\begin{equation} 
\Delta \equiv r^2+a^2-2M r, 
\label{(1.3)}
\end{equation}
\begin{equation}
\Sigma \equiv r^{2}+a^{2}\cos^{2}\theta.
\label{(1.4)}
\end{equation}
For the Kerr metric, there is a unique sensible choice of observers and
tetrads: the locally non-rotating frame for which the observers' world
lines are $r={\rm constant},\theta={\rm constant},\varphi=\omega t
+{\rm constant}$. The corresponding basis of one-forms for such 
an observer is \cite{bard72}
\begin{equation}
e^{(t)}=\sqrt{\Sigma \Delta \over A} \; dt,
\label{(1.5)}
\end{equation}
\begin{equation}
e^{(r)}=\sqrt{\Sigma \over \Delta} \; dr,
\label{(1.6)}
\end{equation}
\begin{equation}
e^{(\theta)}=\sqrt{\Sigma} \; d\theta,
\label{(1.7)}
\end{equation}
\begin{equation}
e^{(\varphi)}=-{2Mar \sin \theta \over \sqrt{\Sigma A}}dt
+\sqrt{A \over \Sigma}\sin \theta \; d\varphi,
\label{(1.8)}
\end{equation}
where the function $A$ is defined as
\begin{equation}
A \equiv (r^{2}+a^{2})^{2}-a^{2}\Delta \sin^{2}\theta.
\label{(1.9)}
\end{equation}
and each basis one-form in (1.5)--(1.8) reads as
\begin{equation}
e^{(i)}=e_{\mu}^{(i)}dx^{\mu}.
\label{(1.10)}
\end{equation}

On the other hand, from the point of view of current developments
in field theory, the axi-symmetry of the Kerr metric makes it
interesting to study how non-commutativity would affect it.
Following Ref. \cite{nico05}, one can assume that non-commutativity of
space-time can be encoded in the commutator of operators
corresponding to space-time coordinates, i.e. (the integer $D$
below is even)
\begin{equation}
\Bigr[x^{\mu},x^{\nu}\Bigr]=i \Lambda^{\mu \nu}, \;
\mu,\nu=1,2,...,D, \label{(1)}
\end{equation}
where the antisymmetric matrix $\Lambda^{\mu \nu}$ is taken to have
block-diagonal form
\begin{equation}
\Lambda^{\mu \nu}={\rm diag}
\Bigr(\Lambda_{1},...,\Lambda_{D/2}\Bigr), \label{(2)}
\end{equation}
with
\begin{equation}
\Lambda_{i}=\Lambda \left(
\begin{array}{cc}
0 & 1 \\
-1 & 0 \\
\end{array}
\right), \; \forall i =1,2,...,D/2, \label{(3)}
\end{equation}
the parameter $\Lambda$ having dimension of length squared and
being constant. The author of Ref. \cite{nico05} solves the Einstein
equations with mass density of a static, spherically symmetric,
smeared particle-like gravitational source as (hereafter we work
in $G=c={\hbar}=1$ units)
\begin{equation}
\rho_{\Lambda}(r)={M\over (4\pi \Lambda)^{3\over 2}}
{\rm e}^{-{r^{2}\over
4\Lambda}}. \label{(4)}
\end{equation}
We here use a different non-commutative prescription 
(which also differs from the work in Refs. 
\cite{cham01,asch05,chai08,mukh08})
to analyze the modification of the Kerr metric.

The plan of the paper is as follows. In Sec. 2 we
apply the first-order Seiberg--Witten map first obtained
by Miao and Zhang \cite{miao10} to the deformed Kerr metric.
The resulting formulae turn out to be 
cumbersome and are, to our knowledge, the first 
systematic evaluation of a first-order Seiberg--Witten
map when one starts from a classical tetrad which describes
the Kerr geometry. Section 3 evaluates the modified efficiency 
of the Penrose process of rotational energy extraction from a
Kerr black hole. Concluding remarks and open problems are
presented in Sec. 4.

\section{$GL(2,C)$ gravity and first-order Seiberg--Witten map}

On considering $SL(2,C)$ gravity in the ordinary space-time, the 
basic physical quantities are the tetrad $e_{\mu}^{a}$ and
spin-connection $\omega_{\mu}^{ab}$, the latter being given by
\begin{equation}
\omega_{\mu}^{ab}={1\over 2}e^{a\nu}
\Bigr(e_{\nu,\mu}^{b}-e_{\mu,\nu}^{b}\Bigr)
-{1\over 2}e^{b\nu}\Bigr(e_{\nu,\mu}^{a}
-e_{\mu,\nu}^{a}\Bigr)
+{1\over 2}e^{a\nu}e^{b \sigma}
\Bigr(e_{\nu,\sigma}^{c}-e_{\sigma,\nu}^{c}\Bigr)e_{c \mu}.
\label{(2.1)}
\end{equation}
With a standard notation, the Latin letters $a,b=0,1,2,3$ are 
Lorentz indices, which are raised and lowered by the Minkowski
metric tensor $\eta_{ab}={\rm diag}(-1,1,1,1)$, while Greek letters
$\mu,\nu=0,1,2,3$ are spacetime indices which are raised and 
lowered by the space-time metric
\begin{equation}
g_{\mu \nu}=e_{\mu}^{a}e_{\nu}^{b}\eta_{ab}
=e_{b \mu} e_{\nu}^{b}.
\label{(2.2)}
\end{equation}
The spin-connection and tetrad can be rewritten as a matrix-valued
one-form with spinor notation, i.e.
\begin{equation}
e=e_{\mu}^{a}\gamma_{a}dx^{\mu}, \;
\omega={1\over 2}\omega_{\mu}^{ab}\sigma_{ab}dx^{\mu}
=\omega_{\mu}dx^{\mu},
\label{(2.3)}
\end{equation}
where $\sigma_{ab}=-{i \over 4}[\gamma_{a},\gamma_{b}]$ are
$SL(2,C)$ generators and $\gamma_{a}$ are the Dirac 
$\gamma$-matrices satisfying the Clifford algebra anticommutation
relations $\left \{\gamma_{a},\gamma_{b} \right \}=2 \eta_{ab}$.

As is well described in \cite{miao10}, \cite{asch09}, the group
$SL(2,C)$ does not close on noncommutative spacetimes and it should
be enlarged to a bigger group by including the additional generators
$1$ and $\gamma_{5}=i \gamma_{0}\gamma_{1}\gamma_{2}\gamma_{3}$, so
that $SL(2,C)$ is enlarged to $GL(2,C)$. The tetrad should then be
extended to include the additional generator $\gamma_{5}\gamma_{a}$
(see the Appendix and Refs. \cite{dewi65}, \cite{espo08}). 
To sum up, the $GL(2,C)$ spin-connection ${\widehat \omega}_{\mu}$
and gauge parameter ${\widehat \psi}$ 
can be decomposed as \cite{miao10}
\begin{equation}
{\widehat \omega}_{\mu}={1\over 2}{\widehat \omega}_{\mu}^{(0)ab}
\; \sigma_{ab}+{\widehat a}_{\mu}^{(1)}
+i{\widehat b}_{\mu 5}^{(1)}\gamma_{5},
\label{(2.4)}
\end{equation}
\begin{equation}
{\widehat \psi}={1\over 2}{\widehat \psi}^{(0)ab} \; 
\sigma_{ab}+{\widehat \psi}^{(1)}
+i {\widehat \psi}_{5}^{(1)}\gamma_{5},
\label{(2.5)}
\end{equation}
while the tetrad is generalized to
\begin{equation}
{\widehat e}_{\mu}={\widehat e}_{\mu}^{(0)a}\gamma_{a}
+{\widehat e}_{\mu 5}^{(1)a} \gamma_{5}\gamma_{a}.
\label{(2.6)}
\end{equation}

The Seiberg--Witten map \cite{seib99} for the tetrad 
${\widehat e}_{\mu}$ is \cite{miao10}
\begin{equation}
{\widehat e}_{\mu}+\delta_{{\widehat \psi}}
{\widehat e}_{\mu}={\widehat e}_{\mu}\Bigr(
e+\delta_{\psi}e,\omega+\delta_{\psi}\omega \Bigr).
\label{(2.7)}
\end{equation}
The solution of Eq. (2.7) up to first order in the non-commutativity
tensor $\Lambda^{\mu \nu}$ reads as \cite{miao10}
\begin{equation}
{\widehat e}_{\mu}^{(0)a}=e_{\mu}^{a},
\label{(2.8)}
\end{equation}
\begin{equation}
{\widehat e}_{\mu 5}^{(1)a}={1\over 4}\Lambda^{\lambda \sigma}
\omega_{\lambda}^{fb}\left(\partial_{\sigma}e_{\mu}^{c}
-{1\over 2}e_{\mu}^{d}\omega_{\sigma}^{ce}\eta_{ed}\right)
\varepsilon_{fbc}^{\; \; \; \; \; a},
\label{(2.9)}
\end{equation}
where $e_{\mu}^{a}$ in (2.8) is the classical tetrad.

In the Kerr geometry the only non-vanishing component of 
$\Lambda^{\mu \nu}$ turns out to be $\Lambda^{23}$
because the metric depends on $(r,\theta)$ only, and hence we find
(with our notation, the superscript $(p,q)$ denotes $p$ derivatives
with respect to $r$ and $q$ derivatives with respect to $\theta$) 
\begin{eqnarray}
&&\hat{e}^{(1) t}_{t 5} = \frac{1}{2\,
{\sqrt{\Sigma (r,\theta)}}\,
    {\left( A(r,\theta )\,\Sigma (r,\theta ) \right) }^
     {\frac{3}{2}}\,{\sqrt{\frac{\Sigma (r,\theta )}
        {\Delta (r,\theta )}}}}\left\{257\,M\,a \,{{\Lambda}^{23}}\,
    \times \right.
    \nonumber \\ 
&& \left( \frac{1}{{\Delta (r,\theta )}^2}\left(r\,
\left( -\left( \Sigma (r,\theta )\,
              \Delta ^{(0,1)}(r,\theta ) \right)  +
           \Delta (r,\theta )\,\Sigma ^{(0,1)}(r,\theta )
           \right)\times \right.\right.
\nonumber \\ 
&& \left. \,\left( 33\,\sin (\theta )\,\Sigma (r,\theta )\,
            A^{(0,1)}(r,\theta ) +
           A(r,\theta )\,\left( -2\,\cos (\theta )\,
               \Sigma (r,\theta ) +
              \sin (\theta )\,\Sigma ^{(0,1)}(r,\theta ) \right)
           \right) \right) \nonumber \\  
&& + \sin (\theta )\,\Sigma^{(1,0)}(r,\theta )\,
(33\,r\,\Sigma (r,\theta )\,A^{(1,0)}(r,\theta ) 
\nonumber \\ 
&& \left. +A(r,\theta )\,(-34\,\Sigma (r,\theta ) +
r\,\Sigma ^{(1,0)}(r,\theta))))\right\},
\label{(2.10)}
\end{eqnarray}
\begin{eqnarray}
&&\hat{e}^{(1) t}_{\varphi 5}= \frac{1}{4\,
    {\sqrt{\frac{A(r,\theta )}{\Sigma (r,\theta )}}}\,
    {\Sigma (r,\theta )}^{\frac{9}{2}}}\left\{-257\,{{\Lambda }^{23}}\,
    {\left( \frac{\Sigma (r,\theta )}{\Delta (r,\theta )} \right)
        }^{\frac{3}{2}}\,\right.\times \nonumber \\&&\left( \sin (\theta )\,
       \Sigma (r,\theta )\,
       \left( \Sigma (r,\theta )\,A^{(0,1)}(r,\theta )\,
          \Delta ^{(0,1)}(r,\theta ) -
         \Delta (r,\theta )\,
          \left( A^{(0,1)}(r,\theta )\,
             \Sigma ^{(0,1)}(r,\theta ) \right.\right.\right.
\nonumber \\&& 
\left.\left.+ \Delta (r,\theta )\,A^{(1,0)}(r,\theta )\,
             \Sigma ^{(1,0)}(r,\theta ) \right)  \right)  +
      A(r,\theta )\,\left( 2\,\cos (\theta )\,
          {\Sigma (r,\theta )}^2\,\Delta ^{(0,1)}(r,\theta ) -
         \Sigma (r,\theta )\, 
          \left( 2\,\cos (\theta )\,
\Delta (r,\theta )\right.\right.\nonumber \\ && \left.\left.\left.\left.+
            \sin (\theta )\,\Delta ^{(0,1)}(r,\theta ) \right) \,
          \Sigma ^{(0,1)}(r,\theta ) +
         \sin (\theta )\,\Delta (r,\theta )\,
          \left( {\Sigma ^{(0,1)}(r,\theta )}^2 +
            \Delta (r,\theta )\,{\Sigma ^{(1,0)}(r,\theta )}^2
            \right)  \right)  \right) \right\},
\label{(2.11)}
\end{eqnarray}
\begin{eqnarray}
&&\hat{e}^{(1) r}_{r 5}= \frac{1}{
    {\Delta (r,\theta )}^2\,
    {\left( A(r,\theta )\,\Sigma (r,\theta ) \right) }^
     {\frac{3}{2}}\,{\sqrt{\frac{\Delta (r,\theta )\,
          \Sigma (r,\theta )}{A(r,\theta )}}}}
\times \nonumber \\ && \left\{4112\,M\,a \,\sin (\theta)\,
    {{{\Lambda }}^{23}}\,{\sqrt{\Sigma (r,\theta )}}\,
    \left( r\,\Sigma (r,\theta )\,A^{(0,1)}(r,\theta )\,
       \Delta ^{(0,1)}(r,\theta ) -
      \Delta (r,\theta )\,\left( r\,A^{(0,1)}(r,\theta )\,
\Sigma ^{(0,1)}(r,\theta ) \right.\right.
\right.\nonumber \\ && +
         \Delta (r,\theta )\,
          \left.\left.\left.\left( -A(r,\theta ) 
+ r\,A^{(1,0)}(r,\theta ) \right)
            \,\Sigma ^{(1,0)}(r,\theta ) \right)  \right) \right\},
\label{(2.12)}
\end{eqnarray}
\begin{eqnarray}
&&\hat{e}^{(1) r}_{\theta 5}= \frac{1}{{\left( A(r,\theta
)\,\Sigma (r,\theta ) \right) }^
     {\frac{3}{2}}\,{\sqrt{\frac{\Delta (r,\theta )\,
          \Sigma (r,\theta )}{A(r,\theta )}}}}\left\{-256\,M\,a \,\sin
(\theta )\,
    {{{\Lambda }}^{23}}\,{\sqrt{\Sigma (r,\theta )}}\times
   \right.\nonumber \\ && \left.\left( A(r,\theta )\,
\Sigma ^{(0,1)}(r,\theta ) -
      r\,\Sigma ^{(0,1)}(r,\theta )\,A^{(1,0)}(r,\theta ) +
      r\,A^{(0,1)}(r,\theta )\,\Sigma ^{(1,0)}(r,\theta ) \right)
    \right\},
\label{(2.13)}
\end{eqnarray}
\begin{eqnarray}
&&\hat{e}^{(1) \theta}_{r 5}= \frac{1}{\Delta (r,\theta )\,
    {\left( A(r,\theta )\,\Sigma (r,\theta ) \right) }^
     {\frac{3}{2}}\,{\sqrt{\frac{\Delta (r,\theta )\,
          \Sigma (r,\theta )}{A(r,\theta )}}}}
\left\{256\,M\,a \,\sin
(\theta )\,
    {{{\Lambda }}^{23}}\,
    {\sqrt{\frac{\Sigma (r,\theta )}{\Delta (r,\theta )}}}\,
    \times \right.\nonumber \\ && \left( A(r,\theta )\,
\left( -\left( \Sigma (r,\theta )\,
            \Delta ^{(0,1)}(r,\theta ) \right)  +
         \Delta (r,\theta )\,\Sigma ^{(0,1)}(r,\theta ) 
\right)\right. \nonumber \\ && +
       r\,\left( \Sigma (r,\theta )\,
          \left( \Delta ^{(0,1)}(r,\theta )\,
             A^{(1,0)}(r,\theta ) -
            A^{(0,1)}(r,\theta )\,\Delta ^{(1,0)}(r,\theta )
            \right) \right. \nonumber \\ && \left.\left.\left.
+ \Delta (r,\theta )\,
          \left( -\left( \Sigma ^{(0,1)}(r,\theta )\,
               A^{(1,0)}(r,\theta ) \right)  +
A^{(0,1)}(r,\theta )\,\Sigma ^{(1,0)}(r,\theta )
            \right)  \right)  \right) \right\},
\label{(2.14)}
\end{eqnarray}
\begin{eqnarray}
&&\hat{e}^{(1) \theta}_{\theta 5}=\frac{1}{\sqrt{\Delta(r,\theta
)}}\hat{e}^{(1) r}_{r 5}=\frac{1}{
    \Delta (r,\theta )\,{\left( A(r,\theta )\,
        \Sigma (r,\theta ) \right) }^{\frac{3}{2}}\,
    {\sqrt{\frac{\Delta (r,\theta )\,\Sigma (r,\theta )}
        {A(r,\theta )}}}}\nonumber \\ && \left\{4112\,M\,a \,\sin
(\theta )\,
    {{{\Lambda }}^{23}}\,
    {\sqrt{\frac{\Sigma (r,\theta )}{\Delta (r,\theta )}}}\,
    \left( r\,\Sigma (r,\theta )\,A^{(0,1)}(r,\theta )\,
       \Delta ^{(0,1)}(r,\theta ) -
      \Delta (r,\theta )\,\left( r\,A^{(0,1)}(r,\theta )\,
          \Sigma ^{(0,1)}(r,\theta )\right.\right.\right.\nonumber \\ &&  +
         \Delta (r,\theta )\,
          \left( -A(r,\theta ) + r\,A^{(1,0)}(r,\theta ) \right)
            \,\Sigma ^{(1,0)}(r,\theta ) \left.\left.\left.\right)  
\right) \right\},
\label{(2.15)}
\end{eqnarray}
\begin{eqnarray}
&&\hat{e}^{(1) \varphi}_{t 5}= \frac{1}{4\,{A(r,\theta )}^2\,
    {\Sigma (r,\theta )}^{\frac{7}{2}}\,
    {\sqrt{\frac{\Delta (r,\theta )\,\Sigma (r,\theta )}
        {A(r,\theta )}}}}\left\{257\,{{\Lambda}^{23}}{\left( 
\frac{\Sigma (r,\theta )}{\Delta (r,\theta )} \right)
        }^{\frac{3}{2}}\,\times \right.\nonumber \\ &&
    \left( {\Sigma (r,\theta )}^3\,
       \Delta ^{(0,1)}(r,\theta )\,
       \left( \Delta (r,\theta )\,A^{(0,1)}(r,\theta ) -
         A(r,\theta )\,\Delta ^{(0,1)}(r,\theta ) \right.\right)  
\nonumber \\ && 
-128\,M^2\,a^{2}r \,{\sin (\theta )}^2\,
       \Delta (r,\theta )\,
       \left( r\,A^{(0,1)}(r,\theta )\,
          \Sigma ^{(0,1)}(r,\theta ) \right.\nonumber \\ && \left.+
         \Delta (r,\theta )\,
          \left( -A(r,\theta ) + r\,A^{(1,0)}(r,\theta ) \right)
            \,\Sigma ^{(1,0)}(r,\theta ) \right)  -
      {\Delta (r,\theta )}^2\,{\Sigma (r,\theta )}^2\,
\left( A^{(0,1)}(r,\theta )\,\Sigma ^{(0,1)}(r,\theta ) 
\right.\nonumber \\ && \left.+
         \left( \Delta (r,\theta )\,A^{(1,0)}(r,\theta ) -
            A(r,\theta )\,\Delta ^{(1,0)}(r,\theta ) \right) \,
          \Sigma ^{(1,0)}(r,\theta ) \right)  \nonumber \\ && +
      \Sigma (r,\theta )\,\left( 128\,M^2\,a^{2}r^{2}\,
          {\sin (\theta )}^2\,A^{(0,1)}(r,\theta )\,
          \Delta ^{(0,1)}(r,\theta ) \right.\nonumber 
\\ && \left.\left.\left.+
         A(r,\theta )\,{\Delta (r,\theta )}^2\,
          \left( {\Sigma ^{(0,1)}(r,\theta )}^2 +
            \Delta (r,\theta )\,{\Sigma ^{(1,0)}(r,\theta )}^2
            \right)  \right)  \right) \right\},
\label{(2.16)}
\end{eqnarray}
\begin{eqnarray}
&&\hat{e}^{(1) \varphi}_{\varphi 5}= \frac{1}{
    {\Sigma (r,\theta )}^{\frac{3}{2}}\,
    {\left( A(r,\theta )\,\Sigma (r,\theta ) \right) }^
     {\frac{3}{2}}\,{\sqrt{\frac{\Delta (r,\theta )\,
          \Sigma (r,\theta )}{A(r,\theta )}}}}\left\{-4112\,M\,a
\,{\sin (\theta )}^2\,{{\Lambda}^{23}}\,
{\sqrt{\frac{A(r,\theta )}{\Sigma (r,\theta )}}}\,
    {\left( \frac{\Sigma (r,\theta )}{\Delta (r,\theta )} \right)
        }^{\frac{3}{2}}\right.\times \nonumber 
\\ && \left( r\,\Sigma (r,\theta )\,
       A^{(0,1)}(r,\theta )\,\Delta ^{(0,1)}(r,\theta ) -
      \Delta (r,\theta )\,\left( r\,A^{(0,1)}(r,\theta )\,
          \Sigma ^{(0,1)}(r,\theta )\right.\right. 
\nonumber \\ &&\left. \left.\left.+
\Delta (r,\theta )\,
\left( -A(r,\theta ) + r\,A^{(1,0)}(r,\theta ) \right)
\,\Sigma ^{(1,0)}(r,\theta ) \right)  \right) \right\}.
\label{(2.17)}
\end{eqnarray}

\section{Modified efficiency of the Penrose process}

In this section we consider the process of rotational energy extraction from
a black hole. In this process proposed by Penrose \cite{penr71}, a
particle falling onto a black hole splits up into two fragments at
some $r > r_+$ where the effective potential $V < 0$, 
and energy can be extracted from a
black hole with an ergosphere. We study the energetics of
Kerr--Newman black hole by the Penrose process using charged
particles in the non-commutative case, by focusing on negative-energy
states. It turns out that the electromagnetic field makes it possible
to extract energy from the black hole. 
In the function $\Delta$ defined in Eq. (1.3) there is an
additional term depending on the charge $Q$ of the black hole, i.e.
\begin{equation}
\Delta=r^2+a^2-2M r + Q^2. 
\label{(3.1)}
\end{equation}
In this space-time there exists an electromagnetic field resulting 
from the presence of charge Q, 
hence the rotation of the black hole gives rise to a magnetic
dipole potential in addition to the usual electrostatic potential.
In the approximation when the metric and the electromagnetic field
are both static and axisymmetric one has two integrals of motion
$E$ and $L$, i.e., the energy and the $\varphi$-component of the
angular momentum per unit of rest mass of the particle, and if the
particle has $p_\theta = 0$ ($p_i$ is the particle's 4-momentum)
in the equatorial plane, it will stay in the plane for all time, i.e.
$p_\theta = 0$ all through the motion. Henceforth the effective
potential for radial motion can be obtained by putting $p_r =
p_\theta = 0$ in the following equation of Ref. \cite{bhat85}:
$$
E= -e A_t - g_{t\varphi}/g_{\varphi\varphi} (L-eA_\varphi) +
(\sqrt{g_{t\varphi}^2-g_{tt}
g_{\varphi\varphi}})/g_{\varphi\varphi}[(L-eA_\varphi)^2
+ g_{\varphi\varphi}(g^{rr}p_r^2 +g^{\theta\theta}p_\theta^2 
+\mu^2)]^{1/2}. 
$$

If one of the fragments has
negative energy (relative to infinity), it will be absorbed by the
black hole while the other fragment will come out, by conservation
of energy, with energy greater than the parent particle. This is
known as the mechanism of energy extraction from the black hole.
In fact, for a test particle of $4$-momentum $p^a = m u^a$, the
energy $E= - p^a \xi_a$ need not be positive in the ergosphere,
hence one can extract energy from a black hole by absorbing a
particle with negative energy \cite{penr02}.

In the case of the Kerr-Newman black hole, the extracted energy
can be given by the rotational and/or the electromagnetic energy
\cite{chri70}. One has to consider the conservation
equations for the 4-momenta of the particles, and one can follow
the recipe for energy extraction given in \cite{bhat85}. At the point
of split, we assume that the 4-momentum is conserved, i.e., $p_1=
p_2+p_3$ where $p_i$ $(i = 1, 2, 3)$ denotes the 4-momentum of the
ith particle. The above relation stands for the following three
relations:
\begin{eqnarray}
E_1=\mu_2 E_2 + \mu_3 E_{3}, \nonumber \\
l_1=\mu_2 l_2+ \mu_3 l_{3}, \nonumber \\
\dot{r}_1=\mu_2 \dot{r}_2+ \mu_3 \dot{r}_{3},
\label{(3.2)}
\end{eqnarray}
where we have set $\mu_1 = 1$. The other conservation relation
follows from the conservation of charge:
\begin{equation}
\lambda_1= \mu_2 \lambda_2+ \mu_3 \lambda_3,
\label{(3.3)}
\end{equation}
where the quantities $\mu_i ,l_i, \lambda_i, E_i, r_i$ refer to
the i-th particle. These relations contain in all eleven
parameters, seven of which are freely specifiable. The choice of
these parameters will be constrained by the requirements that
particle 1 should reach the point of split where $V < 0$ for some
suitable $\lambda$, so that particle 2 can have $E_2 < 0$ and
particle 3 has a runaway orbit.

The most important question in the black hole energetics is the
efficiency of the energy extraction process, for example from a
supermassive black hole, that is one of the many important
parameters of any model in Active Galactic Nuclei. It is therefore
very pertinent to examine how efficient the Penrose process is.

The efficiency $\eta$ is indeed defined as
\begin{eqnarray}
\eta \equiv \frac{{\rm gain}\; {\rm in}\; 
{\rm energy}}{{\rm input}\; {\rm energy}}=
\mu_3\frac{E_3}{E_1} - 1,
\label{(3.4)}
\end{eqnarray}
and it can be calculated in the presence or absence of charge, and/or
electromagnetic interactions. It is known that $\eta_{{\rm max}}\sim
20.7\%$ for the pure extreme Kerr case, in absence of
electromagnetic fields \cite{chan83}. However, there is no upper limit
on $\eta$ in presence of electromagnetic field \cite{part86}.

To evaluate the non-commutativity contribution we are interested in 
formulae which rely upon the tetrad formalism, because the
Seiberg--Witten map is naturally expressed in terms of the tetrad 
rather than the metric. For this purpose, we express $\mu E$ in 
the form \cite{bard72}
$$
\mu E = - p^{(a)}e_{t(a)}= -
p^{(a)}\eta_{ab}e_{t}^{(b)},
$$ 
and we bear in mind that, in the equatorial plane, 
$\theta=\pi/2$, $ p^{(a)}=\mu ( \gamma ,0,
0,\gamma  \,{V_{\varphi }})$, $\Sigma=r^2$, where
$V_{\varphi}$ is the only non-vanishing 3-velocity component.
The classical form of $\mu E$ is
\begin{equation}
\mu E^0=\mu \,\gamma {{{A(r,\theta )}}}^{-1/2}\left( {2\,M\,a
\,{V_{\varphi }}} +
r\,{{{\Delta (r,\theta )}}^{1/2}} \right),
\label{(3.5)}
\end{equation}
while the non-commutative contribution resulting from 
$\hat{e}^{(1) a}_{\mu 5}$ (see (2.6) and (2.9)) is
\begin{eqnarray}
&& \mu E^{NC5}=\frac{{\sqrt{{\Delta (r,\theta )}}}}{4\,r^4\,{\left( \,
A(r,\theta ) \right) }^{\frac{3}{2}}
    }\nonumber \\ && 
\left\{257\,\gamma \,\mu \,{{\Lambda}^{23}}
    \left( \frac{4\,M\,a r^{5}
         \left(-32\,A(r,\theta) +
           33\,r\,A^{(1,0)}(r,\theta ) \right) }{r^2} 
\right.\right.\nonumber \\ && 
+\frac{1}{\,
         {\sqrt{{\,\Delta (r,\theta )}}}}\left(2\,r^2\,{V_{\varphi }}\,
         \left( r\,\left( 128\,M^2\,a^{2} +
              r^2\,\Delta (r,\theta ) \right) \,
            A^{(1,0)}(r,\theta )\right.\right.
\nonumber \\ &&\left.\left.\left.\left. -
           A(r,\theta )\,\left( 128\,M^{2}\,a^{2} +
              2\,r^2\,\Delta (r,\theta ) +
              r^3\,\Delta ^{(1,0)}(r,\theta ) \right)  \right) \right)
      \right) \right\}.
\label{(3.6)}
\end{eqnarray}
To first order in $\Lambda^{23}$, one has therefore
\begin{equation}
\mu E^{(I)}=\mu E^{0} + \mu E^{NC5}.
\label{(3.7)}
\end{equation}
The efficiency $\eta$ defined in (3.4) reduces to
\begin{eqnarray}
\eta= \mu E^{(I)} - 1 ,
\label{(3.8)}
\end{eqnarray}
for $E_1=1$, $\mu_3E_3=\mu E^{(I)}$.
This implies, for the extreme Kerr--Newman black hole %$(a^2 +Q^2 =1)$,
at the horizon ($\Delta =0$) in the absence of charge,  $Q=0$, the
classical efficiency $\eta_{cl}= [\sqrt{2}-1]/2 $ as in
\cite{bhat85}($\mu=1$ for particles), while
\begin{equation}
\eta = [\sqrt{2}-1]/2  - \frac{16448 \,M^2\,
a^{2}{{\Lambda}^{23}}}{\,
    r^3}\,\gamma \,
\,
    {\left( \frac{1}{A(r,\theta )} \right) }^{\frac{3}{2}}\,
    {V_{\varphi }}\,
    \left(-\,r\,\,A^{(1,0)}(r,\theta) +
      A(r,\theta )\, \right).
\label{(3.9)}
\end{equation}
In presence of non-commutativity, the efficiency of a given
energy-extraction event depends on the competition between two
factors in Eq. (3.9), i.e., the standard geometric
term, and the non-commutative one. If the maximum is set at
$0.207$ we have a bound on $\Lambda^{23}$, it must be very small,
and this agrees with our hypothesis.
 
\section{Concluding Remarks}

On the astrophysical side,
several mechanisms have been proposed to account for the power
engine for active galactic nuclei, X-ray binaries, and quasars.
One of the most suggestive is that considered in general
relativity, the Penrose mechanism, that predicts the extraction 
of energy from a rotating black hole.
In the presence of non-commutativity the total energy of a
particle receives a contribution from the non-commutativity parameter in
the efficiency. In this case the extraction of energy from a
non-commutative Kerr--Newman black hole can become more efficient,
although we expect only a tiny gain since the approximate formula
(3.9) for $\eta$ results from a perturbative expansion with a small value
of $\Lambda^{23}$. Our result might apply to the general behaviour 
of negative-energy states and energy extraction process.

We have considered a Kerr black hole immersed in a non-commutative
background (cf. Ref. \cite{kim08}) 
which is perturbative, that is, which does not
appreciably alter the geometrical background although it would
affect significantly the motion of particles. With this assumption
the background geometry has been taken as the one described by the
Kerr metric, perturbatively modified by non-commutativity.
Further developments in sight are as follows.

(i) One has to prove that the Seiberg--Witten map yields a solution
of non-commutative field equations, at least up to some order in the
non-commutativity. For this purpose, we are considering the expansion
of the action functional of Ref. \cite{asch09}, and we hope to be
able to present our results in a separate paper. Recent original work
on non-commutative geometry and the Kerr solution can be found 
in Ref. \cite{bati10}, but the above issue, regarding 
Seiberg--Witten map and non-commutative field equations, remains
unsolved therein as well.

(ii) One can instead make the star product diffeomorphism covariant
\cite{vass09,vass10}, so that the Seiberg--Witten map or twist
become unnecessary.

(iii) One might perform a detailed study of black-hole thermodynamics
in the non-commutative background of Refs. \cite{bane08,bane09} 
pertaining to the applicability of the Penrose 
process in relativistic astrophysics (such applicability was indeed
questioned by Bardeen et al. in Ref. \cite{bard72}).

(iv) Non-commutativity might provide the seed necessary to accelerate 
the fragments to relativistic speeds, without having to assume that
the black hole is immersed in a magnetic field \cite{wagh85}.

\acknowledgments
We are much indebted to Paolo Aschieri for correspondence 
and several scientific discussions, and we are 
grateful to the INFN for financial support. G. Esposito is
grateful to the Dipartimento di Scienze Fisiche 
of Federico II University, Naples, for hospitality
and support; he dedicates the present work to Maria Gabriella.

\appendix
\section{General form of $4 \times 4$ matrices}

It is a simple but non-trivial property that any $4 \times 4$ matrix
can be expressed in the form \cite{dewi65}, \cite{espo08}
\begin{equation}
M_{i}^{\; j}=a \delta_{i}^{\; j}
+b_{\mu}(\gamma^{\mu})_{i}^{\; j}+{1\over 2}c_{\mu \nu}
(\zeta^{\mu \nu})_{i}^{\; j}
+d_{\mu}(\zeta^{\mu})_{i}^{\; j}
+e(\gamma^{5})_{i}^{\; j},
\label{(A1)}
\end{equation}
where
\begin{equation}
(\zeta^{\mu \nu})_{i}^{\; j} \equiv {1\over 2} \left[
(\gamma^{\mu})_{i}^{\; k} (\gamma^{\nu})_{k}^{\; j}
-(\gamma^{\nu})_{i}^{\; k} (\gamma^{\mu})_{k}^{\; j}\right],
\label{(A2)}
\end{equation}
\begin{equation}
(\zeta^{\mu})_{i}^{\; j} \equiv 
(\gamma^{5})_{i}^{\; l} (\gamma^{\mu})_{l}^{\; j}.
\label{(A3)}
\end{equation}


\begin{thebibliography}{}
\bibitem{chri09}
Christodoulou D 2009 {\it The Formation of Black Holes in General
Relativity} (European Mathematical Society) 
\bibitem{kerr63}
Kerr R P 1963 {\it Phys. Rev. Lett.} {\bf 11} 237
\bibitem{wald84}
Wald R M 1984 {\it General Relativity} (Chicago: University of Chicago Press);
Carter B 1972 ``Black hole equilibrium states'' 
in {\it Black Holes}, Les Houches Summer School, eds. De Witt B S and
DeWitt--Morette C (New York: Gordon \& Breach)
\bibitem{hawk73}
Hawking S W and Ellis G F R 1973 {\it The Large Scale Structure of 
Space-Time} (Cambridge: Cambridge University Press)
\bibitem{nico05}
Nicolini P 2005 {\it J. Phys. A: Math. Gen.} {\bf 38} L631
\bibitem{nico06}
Nicolini P, Smailagic A and Spallucci E 2006 {\it Phys. Lett.} 
B {\bf 632} 547
\bibitem{anso07}
Ansoldi S, Nicolini P and Smailagic A 2007 {\it Phys. Lett.}
B {\bf 645} 261
\bibitem{spal09}
Spallucci E, Smailagic A and Nicolini P 2009 {\it Phys. Lett.} B
{\bf 670} 649
\bibitem{nico09}
Nicolini P 2009 {\it Int. J. Mod. Phys.} A {\bf 24} 1229
\bibitem{boye67}
Boyer R H and Lindquist R W 1967 {\it J. Math. Phys.} {\bf 8} 265
\bibitem{dhav00}
Dhavan V, Mirabel I F and Rodriguez L F 2000 {\it Ap. J.}
{\bf 543} 373
\bibitem{toms03}
Tomsik J A and Corbel S 2003 {\it Ap. J.} {\bf 582} 933
\bibitem{blan77}
Blandford R D and Znajek R 1977 {\it Mon. Not. R. Astron. Soc.}
{\bf 179} 433
\bibitem{bard72}
Bardeen J M, Press W H and Teukolsky S A 1972 {\it Ap.J.} {\bf 178} 347
\bibitem{cham01}
Chamseddine A H 2001 {\it Phys. Lett.} B {\bf 504} 33
\bibitem{asch05}
Aschieri P, Blohmann C, Dimitrijevic M, Meyer F, Schupp P 
and Wess J 2005 {\it Class. Quantum Grav.} {\bf 22} 3511
\bibitem{chai08}
Chaichian M, Tureanu A and Zet G 2008 {\it Phys. Lett.} B
{\bf 660} 573
\bibitem{mukh08}
Mukherjee P and Saha A 2008 {\it Phys. Rev.} D {\bf 77} 064014
\bibitem{miao10}
Miao Y G and Zhang S J 2010 arXiv:1004.2118 [hep-th]
\bibitem{asch09}
Aschieri P and Castellani L 2009 {\it JHEP} {\bf 06} 086
\bibitem{dewi65}
DeWitt B S 1965 {\it Dynamical Theory of Groups and Fields}
(New York: Gordon \& Breach)
\bibitem{espo08}
Esposito G 2008 {\it Found. Phys.} {\bf 38} 96
\bibitem{seib99}
Seiberg N and Witten E 1999 {\it JHEP} 99 {\bf 09} 032
\bibitem{penr71}
Penrose R 1971 {\it Nature} {\bf 229} 177
\bibitem{bhat85}
Bhat M, Dhurandhar S and Dadhich N 1985 {\it J. Astrophys. Astr.}
{\bf 6} 85
\bibitem{penr02}
Penrose R 2002 {\it Gen. Rel. Grav.} {\bf 34} 1141
\bibitem{chri70}
Christodoulou D 1970 {\it Phys. Rev. Lett.} {\bf 25} 1596
\bibitem{chan83}
Chandrasekhar S 1983 {\it The Mathematical Theory of Black Holes}
(Oxford: Clarendon Press)
\bibitem{part86}
Parthasarathy S, Wagh S M, Dhurandhar S V and Dadhich N 1986
{\it Ap. J.} {\bf 307} 38
\bibitem{kim08}
Kim H C, Park M I, Rim C and Yee J H 2008 {\it JHEP} {\bf 10} 060
\bibitem{bati10}
Angulo SantaCruz C, Batic D and Nowakowski M 2010 
{\it J. Math. Phys.} {\bf 51} 082504
\bibitem{vass09}
Vassilevich D V 2009 {\it Class. Quantum Grav.} {\bf 26} 145010
\bibitem{vass10}
Vassilevich D V 2010 {\it Class. Quantum Grav.} {\bf 27} 095020
\bibitem{bane08}
Banerjee R, Majhi B R and Samanta S 2008 {\it Phys. Rev.}
D {\bf 77} 124035
\bibitem{bane09}
Banerjee R, Majhi B R and Modak S K 2009 {\it Class. Quantum Grav.}
{\bf 26} 085010
\bibitem{wagh85}
Wagh S M, Dhurandhar S V and Dadhich N 1985 
{\it J. Astrophys. Astr.} {\bf 290} 12
\end{thebibliography}
\end{document}